\documentclass[preprint,floats,aps,epsfig,nofootinbib,amssymb]{revtex4}
\usepackage{mathrsfs}

\usepackage{slashed}
\usepackage{graphicx,color}
\usepackage{epsfig}
\usepackage{subfigure}
\usepackage{epsfig}
\usepackage{dcolumn}
\usepackage{bm}

\begin{document}

\title{Dark Matter, LFV and Neutrino Magnetic Moment in the Radiative Seesaw Model with Triplet Fermion}

\author{Wei Chao}
\email{chaow@physics.wisc.edu}
\ \affiliation{NPAC  Department of Physics, University of Wisconsin-Madison,Madison, WI 53706, USA and Department of Physics, Shanghai Jiao Tong University, Shanghai 200240, China }

\begin{abstract}

In this paper we work in the framework of a radiative seesaw model with triplet fermion $\Sigma$.  Due to the $Z_2$ discrete flavor symmetry, the lightest neutral component of $\Sigma$  is stable and thus can be a dark matter candidate.  Its mass can be solely determined by the dark matter relic abundance, which is bout  $ 2.594 ~{\rm TeV}$.   The model also predict a dark matter-nucleus scattering cross section that would be accessible with future dark matter direct detection searches. We further investigate constraints on the parameter space of the model from the lepton-flavor-violating processes  and neutrino transition magnetic moments, induced by the Yukawa interaction of the $\Sigma$ with the left-handed lepton doublets. .

\end{abstract}

\draft

\maketitle

\section{Introduction}

Although the Standard Model (SM) is in spectacular agreement with the results of most terrestrial experiments, it is certainly fundamentally incomplete. The observation of neutrino oscillations has revealed that neutrinos have non-zero masses and lepton flavors are mixed\cite{pdg}. This is a conclusive evidence of new physics beyond the SM. Besides, precisely cosmological observations have confirmed the existence of non-baryonic cold dark matter: $\Omega_D h^2 =0.1123 \pm 0.0035$\cite{wmap}. Together with the cosmic baryon asymmetry, these important discoveries can not be accommodated in the minimal SM without introducing new ingredients.

In order to generate tiny neutrino masses, one may extend the SM by introducing three right-handed neutrinos having large Majorana masses. Through Yukawa interactions of the right-handed neutrinos with the SM lepton doublets, three active neutrinos then acquire tiny Majorana masses as given by the Type-I seesaw formula: $M_\nu^{} = -M_D^{} M_R^{-1} M_D^T$ \cite{seesawI}, where $M_\nu$ is the mass matrix of light neutrinos, $M_D^{}$ is the neutrino Dirac mass matrix linking the left-handed active neutrinos with the right-handed neutrinos, $M_R^{}$ is the mass matrix of right-handed neutrinos.  Actually, there are three types of tree level seesaw scenarios\cite{seesawI,seesawII,seesawIII} as well as three radiative seesaw scenarios\cite{zee,zeebabu,pma} which  may generate Majorana masses for active neutrinos with the help of dimension-5 Weinberg operator.

Compared with the tree level seesaw models, in which super-heavy seesaw particles have to be introduced, radiative seesaw models have the following two merits: (i) The radiative seesaw models, in which the seesaw particles are usually of the TeV scale and may have large Yukawa interactions with the SM particles, can be tested by the Large Hadron Collider (LHC), while the seesaw particles from the canonical tree level  seesaw models are usually too heavy to be accessible with the LHC;  (ii) The neutral component of seesaw particles in the radiative seesaw models can be stable due to a $Z_2$  discrete flavor symmetry and thus can be the cold dark matter candidate.  This builds a novel connection between  the dark matter and the neutrino physics.

In this paper we focus on a typical radiative seesaw model\cite{made}, which extends the SM with at least two  $Y=0$ fermion triplets $\Sigma$, one inert Higgs doublet $\Phi$,  as well as a $Z_2$ discrete flavor symmetry.   Assuming only $\Sigma$ and $\Phi$ are odd under the $Z_2$ symmetry  and $M_\Sigma < M_\Phi$, the $\Sigma^0$ can be the cold dark matter candidate.  Further assuming the gauge interactions instead of the Yukawa interactions of the $\Sigma$ dominate its  annihilation and co-annihilation in the early universe.  Then the mass of $\Sigma$ as well as its scattering cross section with the nucleon can be solely determined  by the dark matter relic density, as that in the minimal dark matter model\cite{minidm}.  We further study constraints on the parameter space of the model through the lepton-flavor-violating processes and neutrino transition magnetic moments. Our results show that the free parameters of the model can be precisely constrained by these processes.

The paper is organized as follows: Section II is a brief introduction to the model. We study the dark matter phenomenology,  lepton flavor violations and neutrino magnetic moments in section III and IV. Section V is the concluding remarks.

\section{The model}
The model\cite{made} extends the SM with at least two $ Y=0$ Fermion triplets $\Sigma$ having TeV scale Majorana masses, one inert Higgs doublet $\Phi$ and a $Z_2$ discrete flavor symmetry, in which only $\Sigma $ and $\Phi$ are odd and all the other fields are even.   The $Z_2$ invariant Higgs potential is given by \cite{jpma}
\begin{eqnarray}
V= &&-m_1^2 H^\dagger H + m_2^{2} \Phi^\dagger \Phi + {1 \over 2 } \lambda_1^{}  (H^\dagger H)^2  + {1 \over 2 } \lambda_2^{} (\Phi^\dagger \Phi)^2 + \lambda_3^{} (H^\dagger H) (\Phi^\dagger \Phi)\nonumber \\ && + \lambda_4^{} (H^\dagger \Phi) (\Phi^\dagger H) + {1 \over 2 } \lambda_5^{} [(H^\dagger \Phi)^2 + {\rm h.c.}] \; ,
\end{eqnarray}
where  $\lambda_i$ are real parameters. $\langle \Phi^0 \rangle =0$ is required to perserve the exact $Z_2$ symmetry.  The Yukawa interactions for the lepton sector can be given by
\begin{eqnarray}
{-\cal L}_Y = \overline{\ell_L^{i}} Y_{E ij}^{} H E_R^{j} +  \overline{\ell_L^{i}} Y_{ij}^{}\tilde \Phi  \Sigma_R^{j} + {1 \over 2 } M \overline{\Sigma^C} \Sigma + {\rm h.c. } \; ,
\end{eqnarray}
where $M$ is the Majorana mass  matrix of $\Sigma$. There are no Dirac masses linking $\ell_L$  with $\Sigma$ due to the $Z_2$ symmetry. In other words, even though $\Sigma$ have heavy Majorana masses, the canonical seesaw mechanism is not operable.  Therefore the active neutrino masses can  only be generated at the one-loop level with the exchange of $\Sigma^0$ and $\Phi^0$.  The neutrino mass matrix can be written as  \cite{jpma}
\begin{eqnarray}
\left(M_\nu\right)_{ij} \equiv  {\lambda_5^{} v^2 \over 8 \pi^2 } Y_{ik} \zeta_k Y^T_{kj} =   {\lambda_5^{} v^2 \over 8 \pi^2 } \sum_k {Y_{ik} M_k Y^T_{kj} \over M_\Phi^2 - M_k^2 } \left[ 1 - {M_k^2 \over M_\Phi^2 -M_k^2 } \ln {M_\Phi^2 \over M_k^2 }\right]
\end{eqnarray}
where $M_\Phi^2 = (M^2_{ \rm RE {(\Phi_0 })} + M^2 _{ \rm IM {(\Phi_0})})/2$ and
$v=\langle H\rangle$ is the vacuum expectation value of the SM Higgs.

\section{dark matter}

By assuming $M_\Sigma < M_\Phi$,  $\Sigma^0$ is stable and thus can be the cold dark matter candidate. Since the dark matter arises as the thermal relic in the early universe, we may compute its abundance as a function of its mass.  Then we can fit the dark matter mass with the observed dark matter relic density. The final dark matter abundance can be well approximated as \cite{yformula}
\begin{eqnarray}
Y={ n_{\rm DM} \over s } \approx \sqrt{180 \over \pi g_*} {1 \over M_{pl} T_f \langle \sigma_A v \rangle } \; ,
\end{eqnarray}
where $T_f$ is the freeze out temperature, $T_f \sim M/25$,  $g_*$ is the degrees of freedom in thermal equilibrium at the freeze-out temperature, and $s$ is the total entropy.

The relic abundance of $\Sigma^0$ is fully determined by the annihilation and co-annihilation of itself and $\Sigma^\pm$. The most general formulas for the annihilation cross section of the electroweak multiplet dark matter can be found in Ref.\cite{minidm}. Then the thermal average of the annihilation cross section of $\Sigma^0$ can be written as
\begin{eqnarray}
\langle \sigma_A v \rangle \approx { 37 g_2^4 \over 96  M^2 } + \sum_{\alpha \beta} |Y_{\alpha \Sigma } Y_{\beta \Sigma}^* |^2 {6 \over x } {r^2 (1-2r+ 2 r^2 ) \over 24 \pi M^2 } \; ,  \label{sigmav}
\end{eqnarray}
where $g_2$ is the gauge coupling constant of $SU(2)_L$, $r \equiv M^2 /(M_\Phi^2 + M^2 )$ and $x=M/T$. The first term in Eq. (\ref{sigmav}) arises from the gauge interaction and the second term comes from the Yukawa interaction.

\begin{figure}[h!]
\includegraphics[width=8cm]{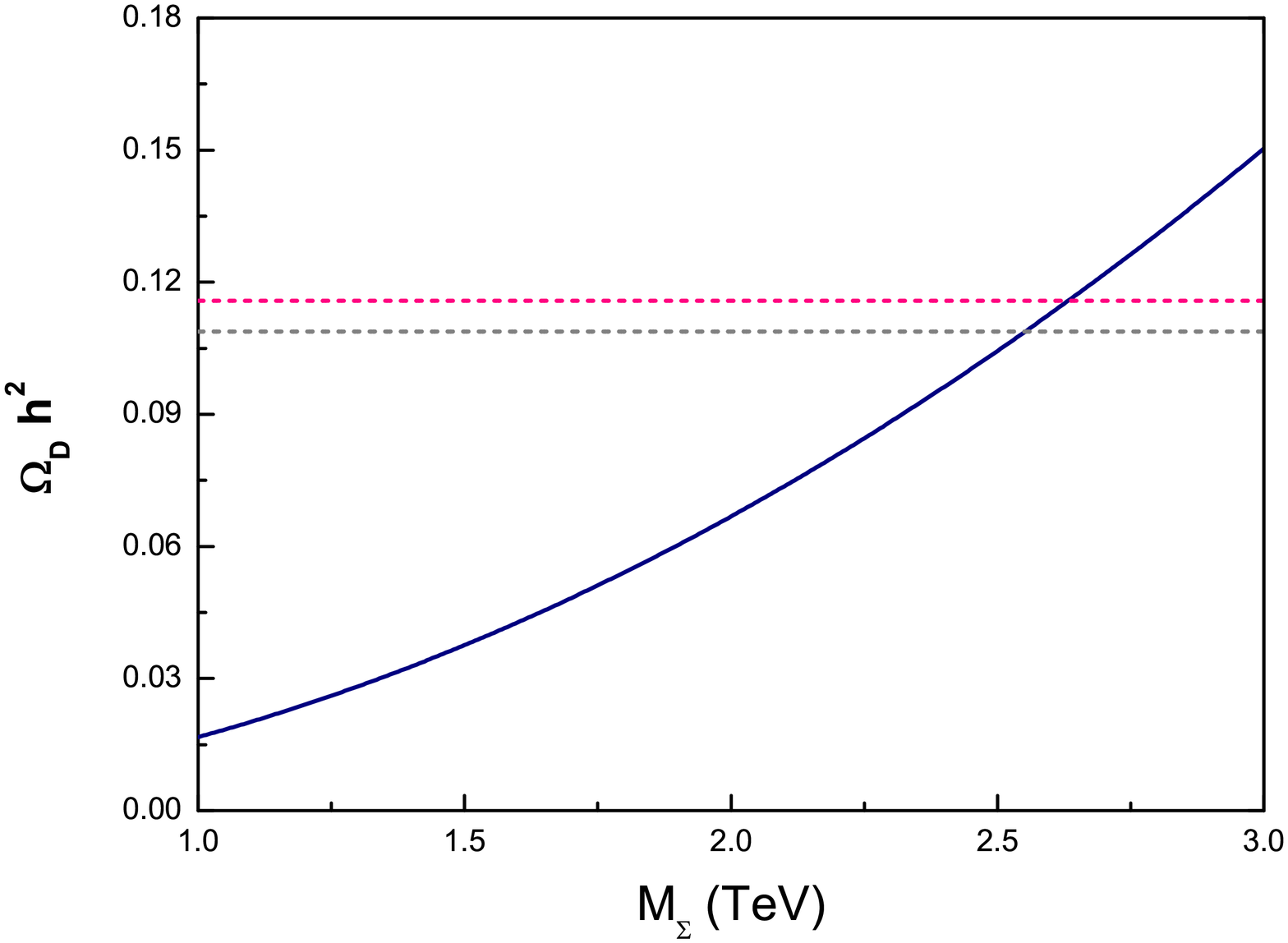}
\includegraphics[width=8cm]{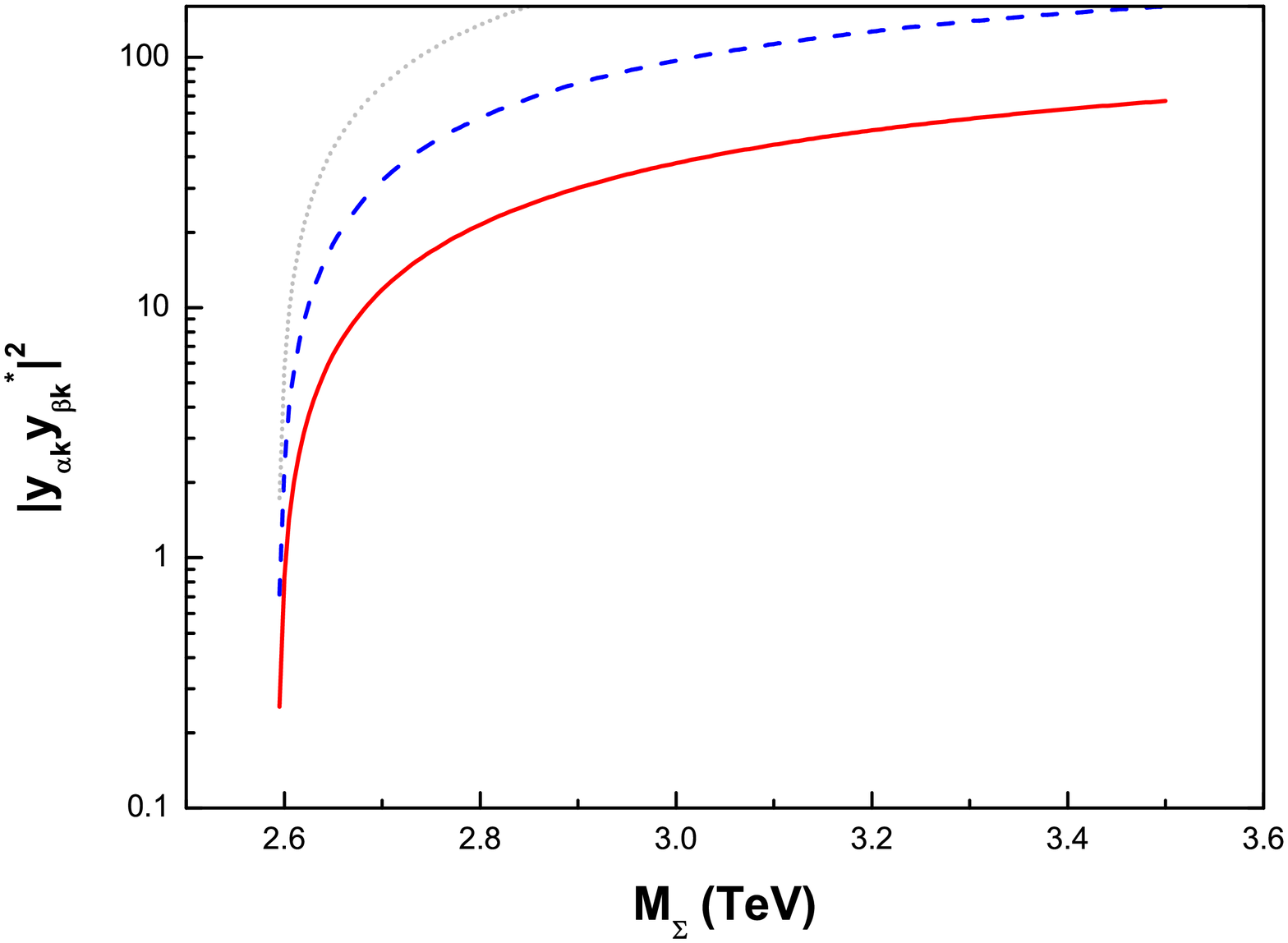}
\caption{ Left panel:  $\Omega_D h^2$ as the function of $M_\Sigma$ for the case of gauge interactions dominating the annihilation of the dark matter. Right panel: $\sum_{\alpha \beta} |Y_{\alpha k} Y_{\beta k}^*|^2 $ as the function of $M_\Sigma$ constrained by the dark matter relic abundance. }
\label{fig:TBrr3}
\end{figure}

The present density of a generic relic is simply given by $\rho_D = m_D  s_0 Y $, where $s_0 = 2889.2 ~{\rm cm^{-3}}$ is the present entropy density. The relic density can finally be expressed in terms of the critical density $\Omega \equiv \rho_D / \rho_C$ , with $\rho_C \equiv 3H^2 /8\pi G_N=1.05\times 10^{-5} h^2 ~{\rm GeV/cm^{3}}$, where $h$ is the dimensionless Hubble parameter.

We assume the first term in Eq. (\ref{sigmav}) dominates the annihilation of $\Sigma$ and plot  in the left panel of  FIG. 1 the $\Omega_D h^2 $ as the function of dark matter mass $M_\Sigma$, where the horizontal band is the measured value of the dark matter relic density at the $95\% $ confidence level.  We may read immediately from the figure that  to generate a correct dark matter abundance $M_\Sigma$ should be  of  $2.594~{\rm TeV}$.  When $M_\Sigma$ get lager, its annihilation cross section to the gauge bosons would be too small to generate the correct relic abundance. Then the second term in Eq. (\ref{sigmav}) may offer the compensation. We plot in the right panel of the Fig. \ref{fig:TBrr3}, $\sum_{\alpha, \beta} |Y_{\alpha k} Y_{\beta k}^* |^2 $ as the function of $M_\Sigma$ constrained by the dark matter relic density.  The solid, dashed and dotted line correspond to $M_\Phi = 4,~6,~8~{\rm TeV}$, separately.  We find that $M_\Sigma$ can't be too larger that $2.59 ~ {\rm TeV}$ even taking into account the contribution of the Yukawa interaction, otherwise the theory would lose perturbativity.

We now study the signatures of the fermion triplet in the dark matter direct detections. According to\cite{minidm}, the  spin-independent cross section of $\Sigma$ scattering off nucleus can be written as
\begin{eqnarray}
\sigma_{SI} (\Sigma^0 N \rightarrow \Sigma^0  N ) = {\pi \alpha_2^4 M_N^4 f^2 \over M_W^2} \left({1\over  M_W^2} + {1 \over M_h^2 } \right)^2 \; ,
\end{eqnarray}
where $M_N$ is the mass of the target nucleus $N$,  $f$ comes from the parameterization of the nucleonic matrix element: $ \langle N | \sum_q \bar q q | N \rangle \equiv f m_N$, with $m_N$ the nucleon mass.  Given this formula, we can calculate the signatures of $\Sigma^0$ in the dark matter direct detection.  By setting $m_H=126 ~{\rm GeV} $\cite{cms,atlas2} and $f=1/3$\cite{fff},  the spin independent cross section is about $\sigma_{SI} =(M_N/1 ~{\rm GeV})^4 \times 1.02 \times 10^{-45} ~{\rm cm}^2 $, which might be accessible with the future Xenon 1 ton or SuperCDMS dark matter direct searches.

\section{LFV and Neutrino Magnetic moment}

In this section, we will study lepton-flavor-violating (LFV) decays, $\mu-e$ conversion in various nuclei and neutrino transition magnetic moments in detail. These processes are induced by the Yukawa interaction of the left-handed lepton doublet with right-handed fermion triplet, which can be written as
\begin{eqnarray}
\Sigma = \left(  \matrix{T^0/\sqrt{2} & T^+ \cr T^- & -T^0/\sqrt{2}}\right) \; .
\end{eqnarray}
The Yukawa interaction can be expressed by the component fields as
\begin{eqnarray}
Y_{ij} \left[ \overline{\nu_i} {T_j^0 \over \sqrt{2} } \Phi^0 - \overline{\nu_i} {T_j^+ } \Phi^- + \overline{ e_{i}} T_j^- \Phi^0 + \overline{e_i} {T^0_j \over \sqrt{2} } \Phi^- \right] + {\rm h.c.} \; , \label{yukawa}
\end{eqnarray}
where the first term is responsible for the origin of tiny neutrino masses with the
help of the radiative seesaw mechanism, the second term may contribute to the neutrino transition magnetic moment, while the third and fourth terms can induce the lepton-flavor-violating decays.  All these processes will put constraint on the parameter space of $Y_{ij}$. We will study them in sequence.

We first begin with charged lepton radiative decays.  The LFV processes in the type-III seesaw mechanism were already investigated in some references\cite{lfviii}. Here we want to stress that LFV  in the radiative seesaw mechanism is different from that in the conventional type-III seesaw model, where there is mixings between the charged lepton and fermion triplet, that may induce flavor changing neutral currents at the tree level.  While for the radiative seesaw case, LFV is only the loop-level effects.

The  most general amplitude  for the $\mu\rightarrow e \gamma$ takes the form
\begin{eqnarray}
T= {e m_\mu \over 16 \pi^2 } \varepsilon_\alpha ^* (q) \bar u_e(p-a) i \sigma_{\alpha \beta} q^\beta \left(  A_L^{} P_L^{} + A_R^{} P_R^{} \right) u_\mu (p) \; ,
\end{eqnarray}
where $p$ and $q$ are the momenta of muon and photon, and the branching ration is given by
\begin{eqnarray}
{\rm BR} (\mu\rightarrow e \gamma) = {3 e^2 \over 64 \pi^2 G_F^2 } \left( |A_L|^2 + |A_R|^2 \right) \left( 1 - {m_e^2 \over m_\mu^2 } \right)^3 \; ,
\end{eqnarray}
with $G_F$ the fermi constant.  For our model, we have $A_L=0$ and
\begin{eqnarray}
A_R^{} = {Y_{ek} Y_{\mu k}^*  \over 4(M_\Phi^2 -M_k^2 )} \left[ 2 {\cal F} \left(  {M_k^2 \over M_\Phi^2 -M_k^2 }\right) + {\cal K} \left( {M_\Phi^2 \over M_\Phi^2 - M_k^2 } \right) \right] \; ,
\end{eqnarray}
where
\begin{eqnarray}
{\cal F} (x) & =& +{1 \over 3 } + {3 \over 2} x + x^2 - x (1+ x)^2 \ln {\left( 1 + x \over x \right)} \; ,\\
{\cal K } (x) &= & - {1 \over 3} + {3 \over 2 } x -x^2 + x(1-x)^2 \ln {\left(1 + x \over x\right)} \; .
\end{eqnarray}

To obtain the $\mu-e$ conversion rates on different nuclei. We shall start with the effective four-Fermion effective operators $e \Gamma_i \mu \bar q \Gamma_i q$ where $\Gamma_i$ is any $4\times4$ Dirac matrices. Following the notation of Ref \cite{kit,yicai}, we have
\begin{eqnarray}
{\cal L}_{\rm eff}&=& -{4 G_F\over \sqrt{2}} \left [m_\mu \bar e \sigma^{\mu\nu} (\tilde A_R P_R + \tilde A_L P_L) \mu F_{\mu\nu}  + h.c.\right ]\nonumber\\
& & -{G_F\over \sqrt{2}} \left [ \bar e (g_{LS}^{(q)}  P_R  + g_{RS}^{(q)} P_L )\mu\, \bar q q  + \bar e (g_{LP}^{(q)}  P_R  + g_{RP}^{(q)} P_L )\mu\, \bar q \gamma_5 q+ h.c.\right ]\nonumber\\
& & - {G_F\over \sqrt{2}} \left [ \bar e (g_{LV}^{(q)} \gamma^\mu P_L  + g_{RV}^{(q)} \gamma^\mu P_R )\mu\, \bar q\gamma_\mu  q  + \bar e (g_{LA}^{(q)}  \gamma^\mu P_L  + g_{RA}^{(q)} \gamma^\mu P_R )\mu\, \bar q \gamma_\mu \gamma_5 q + h.c. \right ]\nonumber\\
&& - {G_F\over \sqrt{2}} \left [ {1\over 2}\bar e (g_{LT}^{(q)} \sigma^{\mu\nu} P_R  + g_{RT}^{(q)} \sigma^{\mu\nu} P_L )\mu\, \bar q\sigma_{\mu\nu}  q + h.c.\right ]\;.
\end{eqnarray}
For our specific model, we get $g_{(L,R) (S,P,A,T) }^{(q)} =0 $, $g_{RV}^{(q)} =0 $, $\tilde A_R^{} = {\sqrt{2 } /( 128\pi^2  G_F) } A_R^{}$, $\tilde A_L = A_R^{} m_e / m_\mu$ and
\begin{eqnarray}
g_{ LV}^{(q)} = Q Y_{ek } Y_{\mu k }^* {  M_W^2 s_W^2 \over 8 \pi^2 (M_\Phi^2 -M_k^2 )}\left[ {\cal G } \left( { M_k^2 \over M_\Phi^2 -M_k^2 }\right)  + {1 \over 3} {\cal U} \left( {M_k^2 \over M_\Phi^2 - M_k^2 } \right) \right]
\end{eqnarray}
where  $s_W =\sin \theta_W$, with $\theta_W$ the Weinberg angle and
\begin{eqnarray}
{\cal G} (x) &=&-x^2 -{5 \over 2 } x - {11 \over 6} + (1 + x)^3 \ln {\left( 1 + x \over x\right)} \; , \\
{\cal U} (x) &=& + x^2 - {1 \over 2 }x  + {1 \over 3 } + x^3 \ln { \left( 1 + x \over x \right)} \; .
\end{eqnarray}
The final branching ration of the $\mu-e$ conversion can be written as\cite{yicai}
\begin{eqnarray}
BR_{\mu \rightarrow e }^A = R^0_{\mu \rightarrow e } (A) \left|1 + {\tilde{g}_{LV}^p V^{p} (A ) \over A_R^{} D(A)}  + {\tilde g _{LV}^n V^n (A) \over A_R^{} D(A)}\right|^2 BR(\mu\rightarrow e \gamma) \; ,
\end{eqnarray}
with
\begin{eqnarray}
\tilde g^p_{LV} &=&2 g_{LV}^{u} + g_{LV}^d \; , \hspace{1cm } \tilde g_{LV}^n =  g_{LV}^{u} +2  g_{LV}^d  \; ,  \nonumber
\end{eqnarray}
where $D(A)$, $V^p(A)$ and $V^n (A)$ are overlap integrals as a function of atomic number.
\begin{figure}[h!]
\includegraphics[width=8cm]{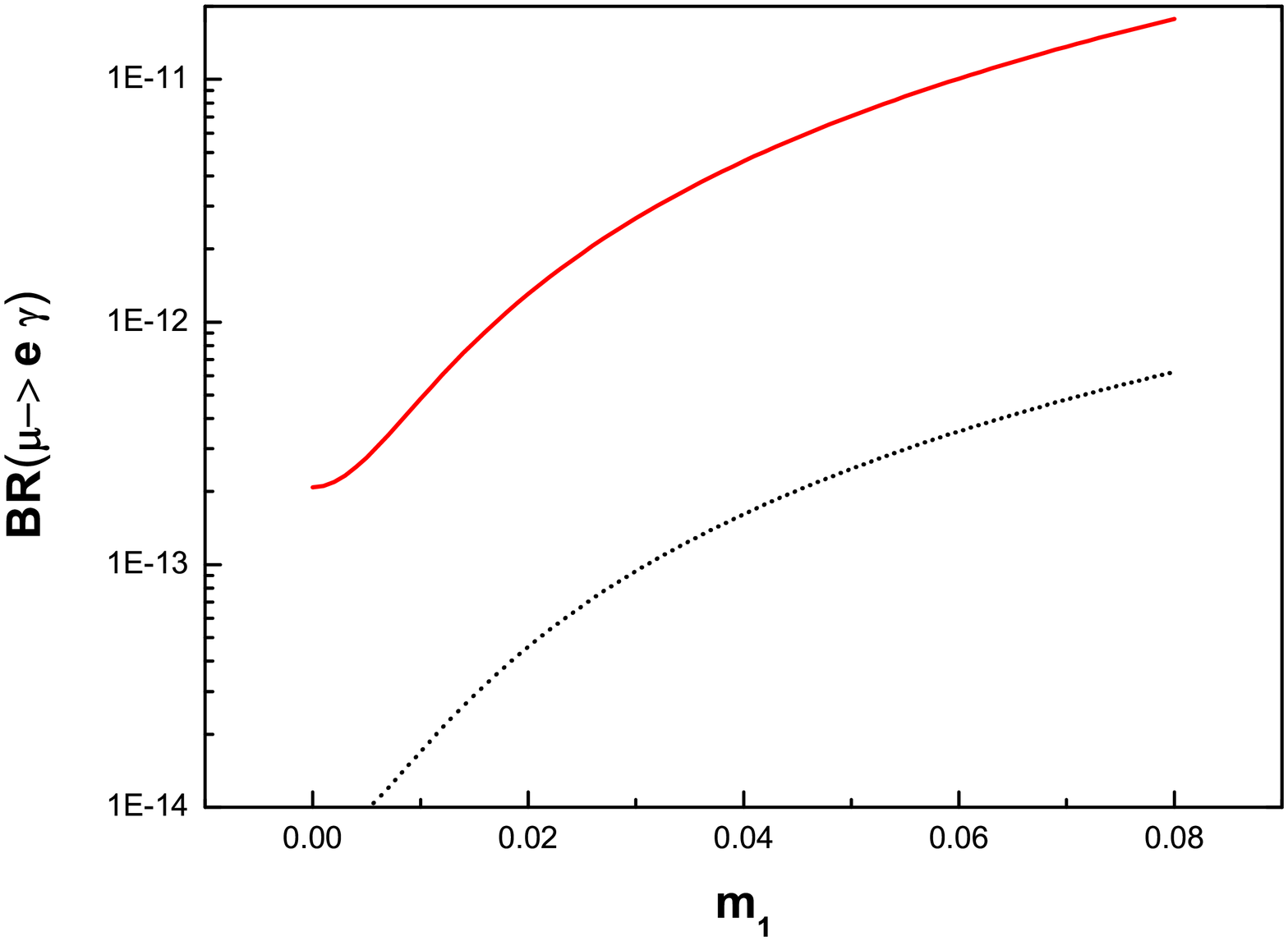}
\includegraphics[width=8cm]{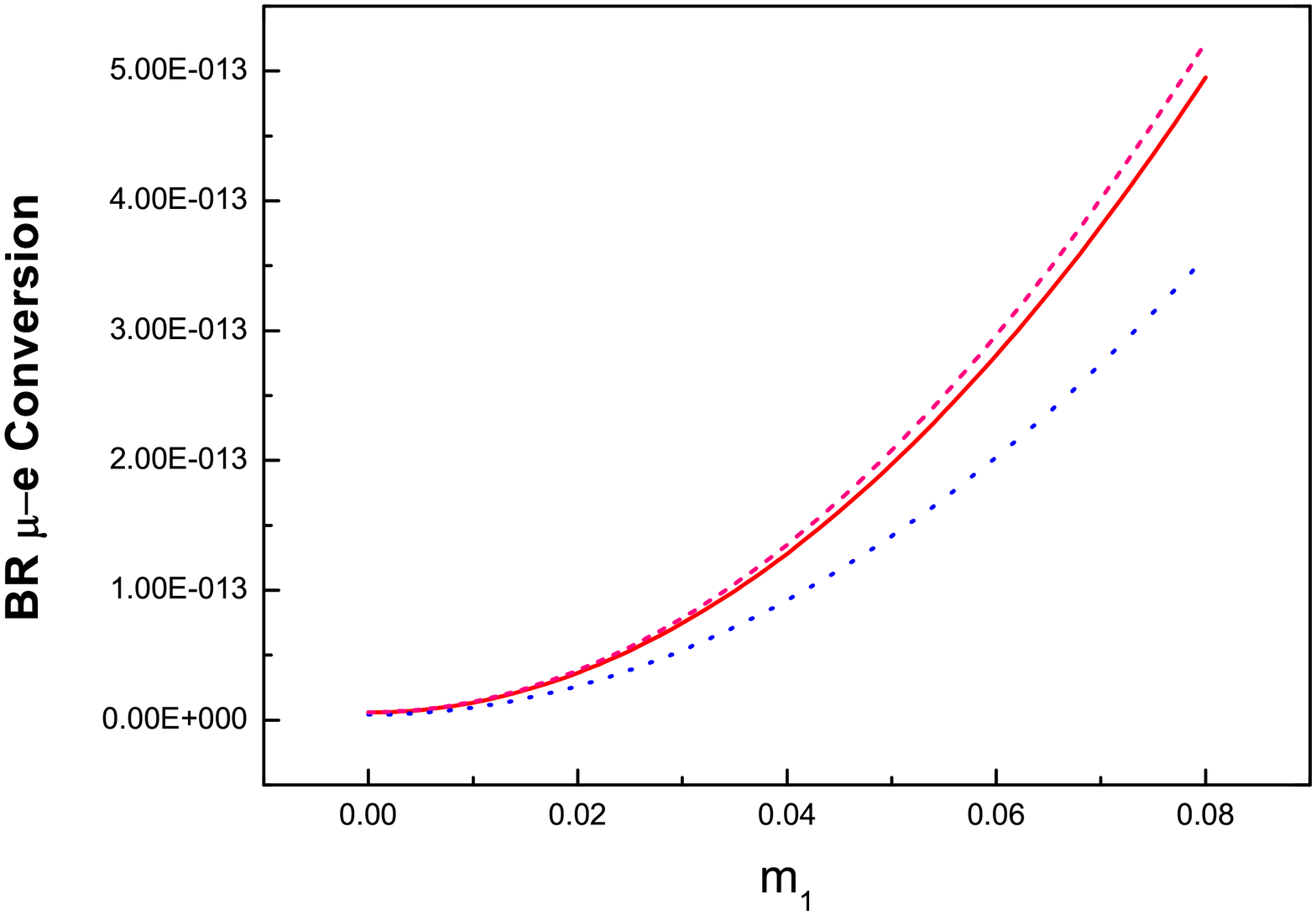}
\caption{ ${\rm BR}(\mu \to e \gamma)$ (left panel) and ${\rm BR} ~(\mu-e)$ conversion in various nuclei (right panel) as the function of $m_1$, assuming light neutrinos in the normal mass hierarchy}
\label{fig:TBrr4}
\end{figure}

\begin{figure}[h!]
\includegraphics[width=10cm]{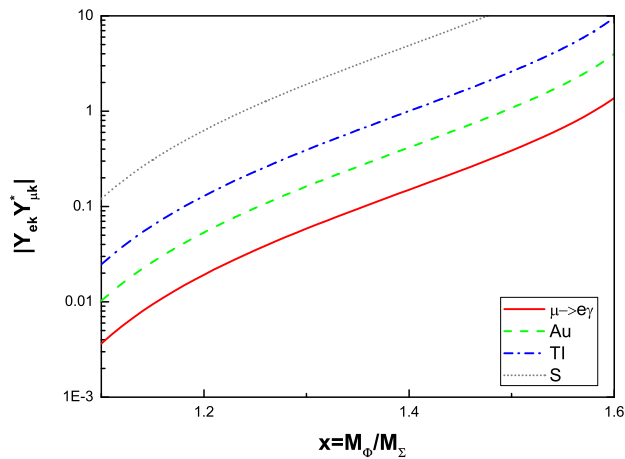}
\caption{ $|Y_{e2} Y_{\mu 2}^*|$ as the function of $x=M_\Phi/M_2$ constrained by the current upper limit of various lepton-flavor-violating processes. }
\label{fig:TBrr5}
\end{figure}



To carry out numerical analysis, we need to parameterize the Yukawa coupling matrix $Y$ with neutrino masses and lepton mixing angles.  A general solution for $Y$ can be written as
\begin{eqnarray}
Y= \left( \lambda_5 v^2 \over 8 \pi^2  \right)^{-1/2} V_{\rm PMNS} \hat M_\nu^{1/2} O \hat\zeta^{1/2}
\end{eqnarray}
Here  $\hat M_\nu $ and $\hat \zeta$ are diagonal matrices with $\hat M_\nu ={\rm diag} \{ m_1, m_2, m_3 \}$ and $\hat\zeta ={\rm diag} \{ \zeta_1, \zeta_2, \zeta_3 \} $, where $\zeta_k$ is defined in Eq. (3). $V_{\rm PMNS}$ is the Pontecorvo-Maki-Nakagawa-Sakata mixing matrix.  $O$ satisfies $O^TO =1$ and can be complex in general.

As an illustration, we take $\lambda=10^{-8}$, $O=I$
, $M_2\ll M_{1,3} $,  and $V_{\rm PMNS}$ to be the tri-bimaximal form\cite{tribimax}. We plot in the left panel of FIG. 2   $ {\rm BR} (\mu \to e \gamma)$ as the function of $m_1$, assuming the light neutrinos in a normal mass hierarchy.  The solid and dotted lines correspond to $x(\equiv M_\Phi/ M_2) = 1.2 ~{\rm and} ~1.5$ separately. We find that the branching will reach its experimental upper limit as $m_1$  gets larger than $0.08$ ${\rm eV}$ for the $x=1.2$ case.   We plot in the right panel of the FIG. 2  ${\rm BR}(\mu - e  ~{\rm conversion} )$ as the function of $m_1$, assuming the light neutrinos in a normal mass hierarchy and $x=1.2$.  The solid, short-dashed and dotted lines correspond to the case of $\mu -e$ conversion in ${\rm Au}$, ${\rm Ti}$ and ${\rm S}$, separately.  The current upper limit of $\mu \to e$ conversion in various nuclei are $7.0\times 10^{-11}$ in ${\rm S}$, $4.3\times 10^{-12}$ in ${\rm Ti}$ and $7.0\times 10^{-13}$ in ${\rm Au}$\cite{kit}. We may conclude from the figure that  the LFV in  ${\rm Ti}$ is more sensitive to  the $m_1$, but the LFV in ${\rm Au}$ puts more stringent constraint.  For comparing, we plot in FIG. 3, $|Y_{e2} Y_{\mu 2}^*|$ as the function of the $x(\equiv M_\Phi/M_2)$ constrained by the current upper limit of various lepton flavor violation processes.  The solid, dashed, dot-dashed and dotted lines correspond to $\mu \to e \gamma$ and $\mu -e $ conversion in ${\rm Au, ~Ti, ~and ~S}$ separately.  The conclusion is $\mu \to e \gamma$ puts the strongest constraint taking into account the current results.  On the other hand, sensitivities of prospective future $\mu-e$ conversion searches\cite{mutoe} will exceed that of the MEG \cite{meg} experiments.
\begin{figure}[h!]
\includegraphics[width=8cm]{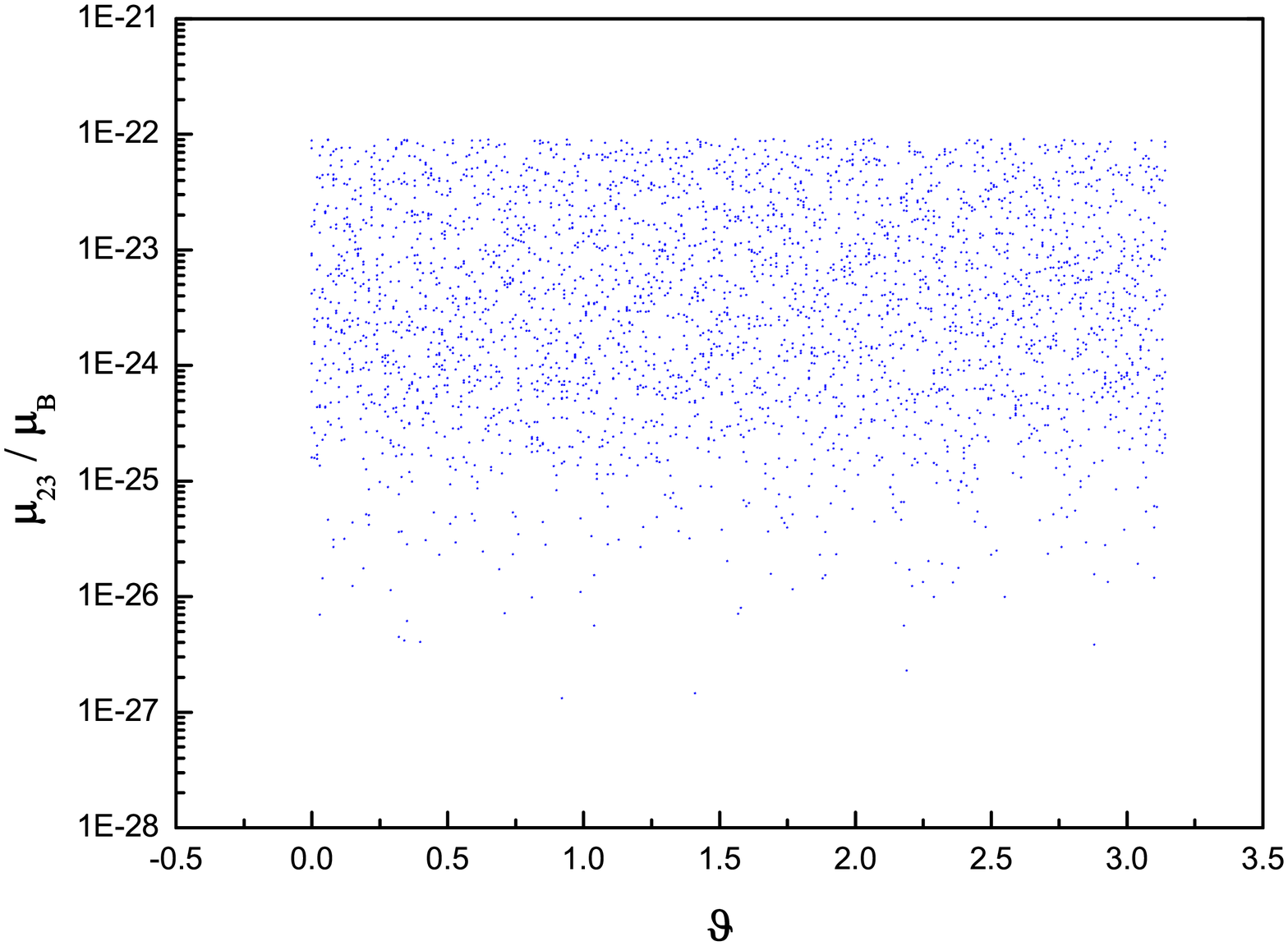}
\includegraphics[width=8cm]{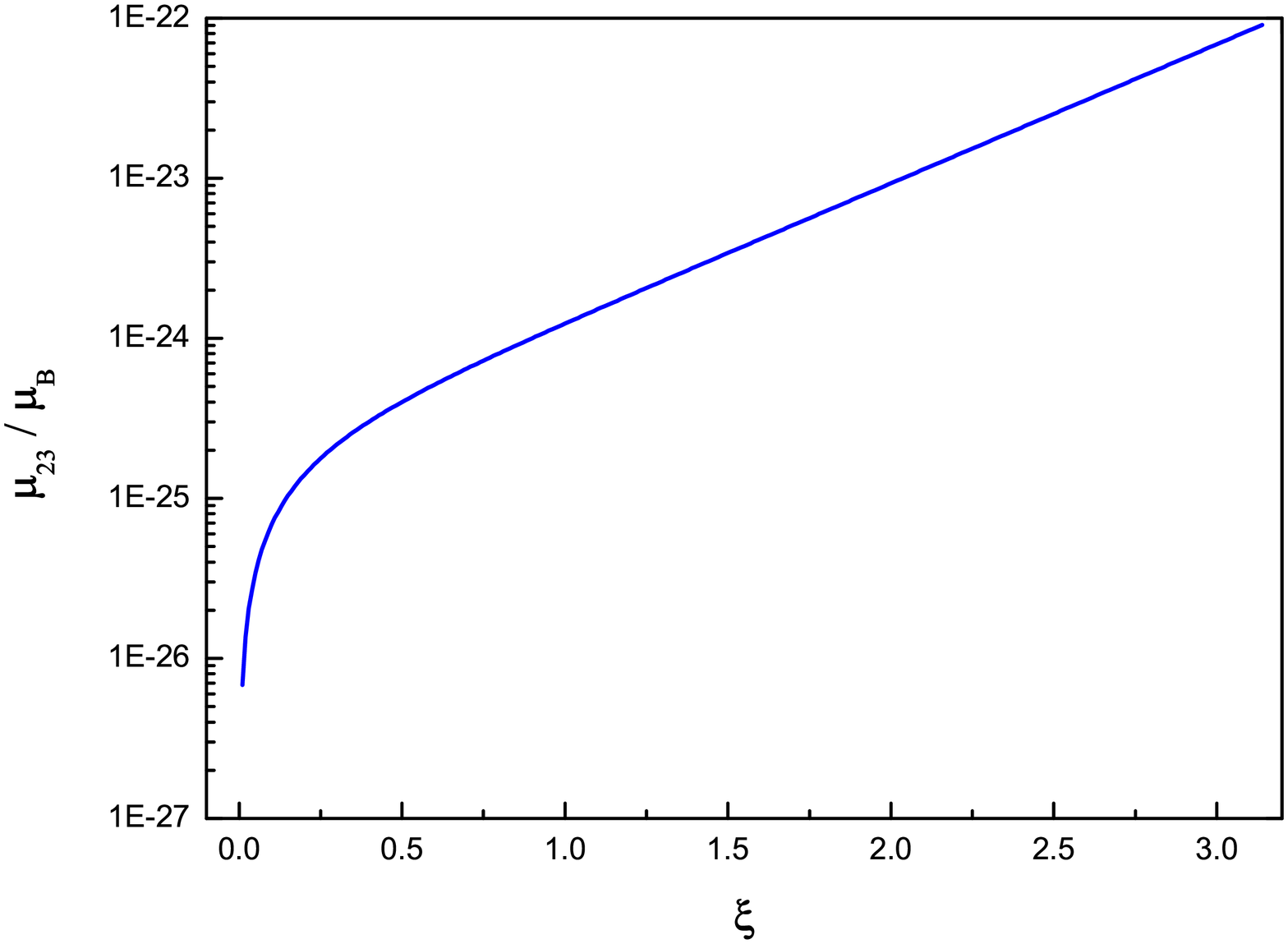}
\caption{ Left panel: ${\mu_{23}/\mu_B}$  as the function of $\vartheta$ by varying $\xi$ in the range $[0,~\pi ]$. Right panel: $\mu_{23}/\mu_B$ as the function of $\xi$ by setting $\vartheta=\pi/3$.  }
\label{xxx}
\end{figure}

Since active neutrinos have non-zero masse, they may have electromagnetic dipole moments, which might be in general sensitive to the unitarity violation of the lepton mixing matrix\cite{xingzhou}.  For the radiative seesaw model we are interested in, the PMNS matrix is unitary, but the  TeV scale triplet fermion may lead to a non-zero electromagnetic form factors at the one-loop level:
\begin{eqnarray}
{e} \bar \nu_i^{} i \sigma_{\alpha \beta } F^{\alpha \beta}\left\{   i {\rm Im}[ \mathcal{A}](m_j  + m_i ) + {\rm Re}[ \mathcal{A}] (m_j  - m_i ) \gamma_5   \right\} \nu_j^{} \; , \label{neuform}
\end{eqnarray}
where
\begin{eqnarray}
\mathcal{A} = -{i\over 32 \pi^2 }\hat Y_{ik} {1 \over M_\Phi^2 -M_k^2 } \left\{{\cal F } \left( {M_k^2 \over M_\Phi^2 -M_k^2 }\right)+  {\cal K} \left( {M_\Phi^2 \over M_\Phi^2 -M_k^2 }\right) \right \}\hat Y^\dagger_{kj }  \; ,
\end{eqnarray}
with
\begin{eqnarray}
\hat Y= \left( \lambda_5 v^2 \over 8 \pi^2  \right)^{-1/2}  \hat M_\nu^{1/2} O \hat\zeta^{1/2}\; .
\end{eqnarray}
We may examine that the magnetic moment of neutrinos is exactly zero due to its Majorana nature, they  have only transition magnetic moments.  Assuming there are only two generation  triplet fermions and the light neutrinos in a normal mass hierarchy, the ${O}$ can be given as
\begin{eqnarray}
O=\left(  \matrix{ 0 & 0 \cr \cos[\vartheta+ i \xi] & -\sin[\vartheta+ i \xi] \cr \sin [\vartheta+ i \xi] & ~~\cos[\vartheta+ i \xi] }\right) \; .
\end{eqnarray}
We plot in the left panel of the FIG. 4, neutrino transition magnetic moment $\mu_{23}$ as the function of $\vartheta$ by varying $\xi$ in the range $[0 , \pi]$ and in the right panel of the FIG. 4, $\mu_{23}$ as the function of $\xi$ by setting $\vartheta = \pi/3$.  We find that the neutrino  magnetic moment is of the order $10^{-23} \mu_B$, which is much smaller than the current experimental upper bound, $\mu_\nu <0.32\times 10^{-10} \mu_B$ at the $90\%$ CL\cite{neumag}.

Given the neutrino form factors in Eq. (\ref{neuform}), we may also estimate the decay rate of active neutrinos, which may contribute to the cosmic infrared background in the Universe\cite{neudecay}.  The neutrino decay rates is given by
\begin{eqnarray}
\Gamma (\nu_j \to \nu_i \gamma) ={\alpha \over 2 m_j^3 } (m_j^2 -m_i^2 )^3 \left(  |{\rm Im} [\mathcal{A}]|^2 + |{\rm Re} [\mathcal{A}]|^2 \right) \; .
\end{eqnarray}
where $\alpha$ is the fine-structure constant. We roughly estimate the decay rate of the process $\nu_3 \to \nu_2 \gamma$, which is of the order ${\cal O}(10^{-69}) $ GeV by setting $M_\Phi= 1.5 M_1$ and $\vartheta=\pi/3$.

\section{Conclusion}

In this paper we work in the framework of the radiative seesaw mechanism with fermion triplet and $Z_2$ discrete flavor symmetry. Assuming $M_\Sigma< M_\Phi$, the lightest neutral component of the fermion triplet can be the dark matter candidate. Further assuming it annihilates mainly through the gauge interaction, then its mass is precisely determined by the dark matter relic abundance, which is about $2.594 ~{\rm TeV}$.  We calculated its scattering cross section with nuclei. Lepton flavor violations as well as neutrino transition magnetic moments were also studied. Our results shows that neutrino magnetic moments is too small compared with the experimental value, while lepton flavor violating processes can be accessible with the current and (or) future LFV searches.

\begin{acknowledgments}
This work was supported in part by the Wisconsin Alumni Research Foundation.

\end{acknowledgments}


\begin{thebibliography}{99}

\bibitem{pdg}

K. Nakamura {\it et al}., (Particle Data Group), J. Phys. G {\bf
37}, 075021 (2010).


\bibitem{wmap}

E. Komatsu, {\it et al}., arXiv:1001.4538[astro-ph.CO].




\bibitem{seesawI}
P.~Minkowski,
  Phys.\ Lett.\ B {\bf 67}, 421 (1977);
  T.~Yanagida, in {\it Workshop on Unified Theories}, KEK report 79-18 p.95 (1979);
  M.~Gell-Mann, P.~Ramond, R.~Slansky,
  in {\it Supergravity} (North Holland, Amsterdam, 1979)
  eds. P.~van~Nieuwenhuizen, D.~Freedman, p.315;
  S.~L.~Glashow, in {\it 1979 Cargese Summer Institute on Quarks and Leptons} (Plenum Press,
  New York, 1980) eds. M.~Levy, J.-L.~Basdevant, D.~Speiser, J.~Weyers, R.~Gastmans and M.~Jacobs,
  p.687;
  R.~Barbieri, D.~V.~Nanopoulos, G.~Morchio and F.~Strocchi,
  Phys.\ Lett.\ B {\bf 90}, 91 (1980);
  R.~N.~Mohapatra and G.~Senjanovic,
  Phys.\ Rev.\ Lett.\  {\bf 44}, 912 (1980);
  G.~Lazarides, Q.~Shafi and C.~Wetterich,
  Nucl.\ Phys.\  B {\bf 181}, 287 (1981).

\bibitem{seesawII}
W.~Konetschny and W.~Kummer,
  Phys.\ Lett.\  B {\bf 70}, 433 (1977);
%
 T.~P.~Cheng and L.~F.~Li,
  Phys.\ Rev.\  D {\bf 22}, 2860 (1980);
%
 G.~Lazarides, Q.~Shafi and C.~Wetterich,
 Nucl.\ Phys.\  B {\bf 181}, 287 (1981);
%
 J.~Schechter and J.~W.~F.~Valle,
  Phys.\ Rev.\  D {\bf 22}, 2227 (1980);
%
 R.~N.~Mohapatra and G.~Senjanovic,
  Phys.\ Rev.\  D {\bf 23}, 165 (1981).


\bibitem{seesawIII}
R.~Foot, H.~Lew, X.~G.~He and G.~C.~Joshi,
  Z.\ Phys.\  C {\bf 44}, 441 (1989).

\bibitem{zee}

A. Zee, Phys. Lett. B {\bf 93}, 389 (1980), Erratum-ibid. B {\bf 95}, 461 (1980).

\bibitem{zeebabu}
A. Zee, Phys. Lett. B {\bf 161}, 141 (1985); A. Zee, Nucl. Phys. B {\bf 264}, 99 (1986); K. S. Babu, Phys. Lett. B {\bf 203}, 132 (1988).

\bibitem{pma}
E. Ma and U. Sarkar, Phys. Rev. Lett. {\bf 80}, 5716 (1998).




\bibitem{made}

E. Ma and D. Suematsu, Mod.Phys. Lett. A {\bf 24}, 583 (2009).

\bibitem{jpma}
E. Ma, Phys. Rev. D {\bf 73}, 077301(2006);
J. Kubo, E. Ma, D. Suematsu, Phys. Lett. B {\bf 642}, 18 (2006); T. Li, W. Chao, nucl. Phys. B {\bf  843}, 396 (2011).

\bibitem{yformula}

E. W. Kolb, M. S. Truner, The early Universe, Addison-Wesley, Reading, MA, 1993; M. Srednicki, R. Watkins, K. A. Olive, Nucl. Phys. B {\bf 310}, 693 (1988); P. Gondolo, G. Gelmini, Nucl. Phys. B {\bf 360}, 145 (1991).


\bibitem{minidm}

M. Cirelli, N. Fornengo and A. Strumia, Nucl. Phys. B {\bf 753}, 178 (2006).

\bibitem{cms}

CMS Collaboration, arXiv:1202.1487; 1202.1416; 1202.1488; 1202.1489.

\bibitem{atlas2}

ATLAS Collaboration, arXiv:1202.1414; 1202.1415.

\bibitem{fff}

M. Dress, M. Bojiri, Phys. Rev. D {\bf 48}, 3483 (1993).

\bibitem{kit}
R.~Kitano, M.~Koike and Y.~Okada, Phys.\ Rev.\  D {\bf 66}, 096002 (2002);[Erratum-ibid.\  D {\bf 76}, 059902 (2007)].



\bibitem{tribimax}

P. F. Harrison, D. H. Perkins, W. G. Scott, Phys. Lett. B {\bf 530}, 167 (2002); Z. Z. Xing, Phys. Lett. B {\bf 533}, 85 (2002); X. G. He and A. Zee, Phys. Lett. B {\bf 560}, 87 (2003);


\bibitem{mutoe}

J. P. Miller [Mu2E collaboration], {\it Proposal to search for $\mu^- N \rightarrow e^- N $ with a singlet event sensitivity below $10^{16}$}; Y. Kuno, {\it et al}., [COMET collaboration], {\it An experimental search for lepton flavor violating $\mu-e$ conversion at sensitivity of $10^{-16}$ with a slow-extracted bunched beam}; Y. Kuno {\it et al.,} [PRISM/PRIME group], {\it An experimental search for a $\mu-e$ conversion at sensitivity of $10^{-18}$ with a highly intense muon source: PRISM }.

\bibitem{meg}

R. Sawada, (MEG Collaboration), PoS (IHEP 2010) 263.


\bibitem{xingzhou}

Z. Z. Xing and Y. L. Zhou, arXiv:1201.2543.

\bibitem{neudecay}
A. Mirizzi, D. Montanino and P. Serpico, Phys. Rev. D {\bf 76}, 053007 (2007); S. Matsuura {\it et al}., Astrophys. J. {\bf 737}, 2 (2011); S. H. Kim, {\it et al}., arXiv:1112.4568.

\bibitem{neumag}

A. G. Beda, {\it et al}., Phys. Part. Nucl. Lett. {\bf 7}, 406 (2010).

\bibitem{yicai}

Y. Cai, X. G. He, M. Ramsey-Musolf, L. H. Tsai, JHEP {\bf 1112}, 054 (2011).

\bibitem{lfviii}

A. Abada, C. Biggio, F. Bonnet, M. B. Gavela and T. Hambye, Phys. Rev. D {\bf 78}, 033007 (2008); JHEP {\bf 12}, 061 (2007).

\end{thebibliography}
\end{document}